\documentclass{iau} 
\usepackage{graphicx}

\newcounter{qub}

\newcommand{\sunn}{$_{\odot}$}

\title[Very gas-rich XMP blue void dwarfs]
{Very gas-rich extremely metal-poor blue void dwarfs}

\author[S.~Pustilnik, Y.~Perepelitsyna, A.~Kniazev, E.~Egorova, J.~Chengalur]
{Simon A. Pustilnik$^1$, Yulia A.~Perepelitsyna$^1$, Alexei Y.~Kniazev$^{2,3}$,
 Evgeniya S.~Egorova$^3$ \and Jayaram N.~Chengalur$^4$}

\affiliation{$^1$Special Astrophysical Observatory of RAS, \\ 369167,
Nizhnij Arkhyz, Karachai-Circessia, Russia, email: {\tt sap@sao.ru, jlyamina@yandex.ru}\\
[\affilskip]
$^2$South African Astronomical Observatory, \\
1 Observatory road, Observatory, 7935 Cape Town, South Africa
email: {\tt akniazev@saao.ac.sa} \\
[\affilskip]
$^3$Sternberg Astronomical Institute of Moscow State University, \\
Universitetsky Pr. 13, Moscow, Russia, email: {\tt eshaldenkova@gmail.com} \\
[\affilskip]
$^4$National Centre for Radio Astronomy (TIFR), \\ Postbag 3,  Ganeshkind,
Pune 411007, India, email: {\tt chengalur@ncra.tifr.res.in}}

\pubyear{2019}
\volume{344}
\jname{Dwarf Galaxies: From the Deep Universe to the Present}
\editors{K.~McQuinn, S.~Stierwalt, eds.}
\begin{document}

\maketitle

\begin{abstract}
Half-dozen of extreme representatives of void dwarf galaxy population were
found in our study of evolutionary status of a hundred galaxies in the nearby
Lynx-Cancer void. They are very gas-rich, extremely low-metallicity
[$7.0 < 12+\log(O/H)< \sim7.3$] objects, with blue colours of outer parts.
The colours indicate the ages of the oldest visible stellar population of
one to a few Gyr. They all are intrinsically faint, mostly Low Surface
Brightness dwarfs, with $M_{\rm B}$ range of --9.5$^m$ to -14$^m$. Thus,
their finding is a subject of the severe observational selection. The recent
advancement in search for such objects in other nearby voids resulted in
doubled their total number. We summarize all available data on this group
of unusual void dwarf galaxies and discuss them in the general context of
very low metallicity galaxies and their possible formation and
evolutionary scenarios.
\end{abstract}

\firstsection 
\section{Introduction}

The low-metallicity galaxies with gas metallicity $Z < Z$\sunn/10 (known
number more than 350) remain very rare objects. Even more so are XMP
galaxies with $Z < Z$\sunn/20 (or $12+\log$(O/H) $ \lesssim 7.38$, about
50 known galaxies, \cite[Guseva et al. (2017)]{Guseva2017}).
XMP galaxies are important as the best local proxies for forming galaxies in
the early Universe. Most of known low-metallicity galaxies are found via
optical spectroscopy of star-forming galaxies. There is a problem of dearth of
low-metallicity dwarfs (\cite[Sanchez Almeida et al. (2017)]{Sanchez2017}.
One of the exits is the existence of many quiscent, LSB dwarfs with low or
subtle SFR, missed by optical redshift surveys. The alternative means, such
as A) blind wide-angle HI surveys (ALFALFA,
\cite[Haynes et al. (2018)]{Haynes2018}),
B) colour-morphological selection of blue dIr
(e.g., \cite[James et al. (2017)]{James2017},
\cite[Hsyu et al. (2018)]{Hsyu2018})
 and C) the unbiased study of all galaxies in the nearby voids, open
additional channels to identify 'quiscent' XMP dwarfs.
We describe and discuss the most unusual XMP dwarfs found in the
nearby voids.

\begin{figure*}[b]
\begin{center}
 \includegraphics[width=10cm,angle=-90,clip=]{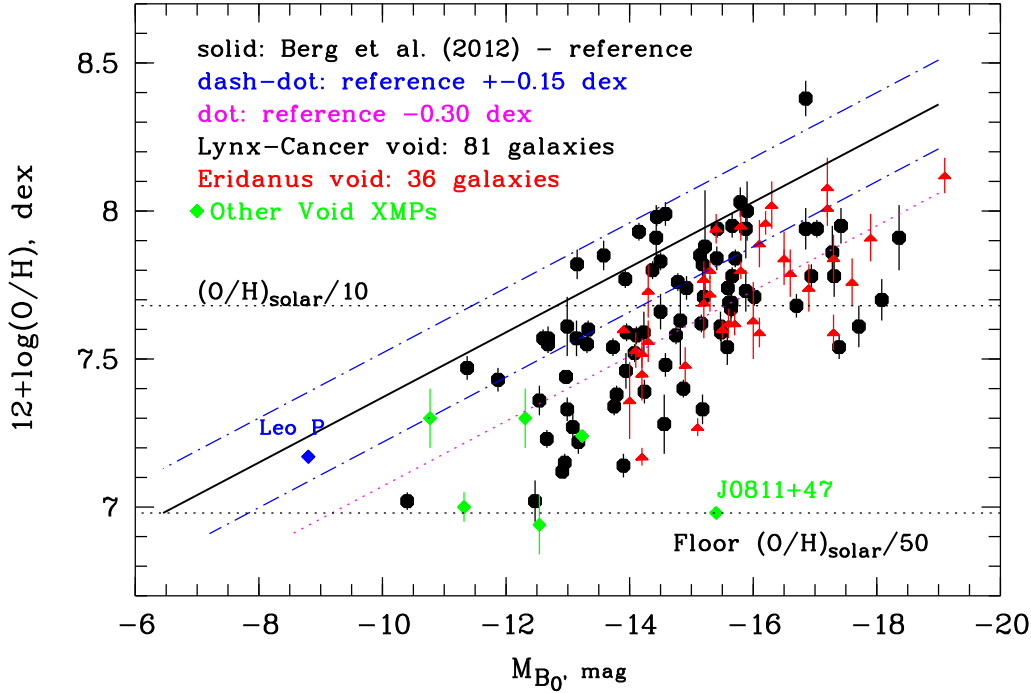}
 \caption{Nearby Void dwarfs on plot '$12+\log$(O/H) versus $M_{\rm B}$' in
comparison with the linear regression (solid) for the reference sample of
late-type galaxies in the Local Volume from
\cite[Berg et al. (2012)]{Berg2012}. Two dashed-dot lines show a one sigma
scatter of the reference sample. Most of void galaxy data are from
\cite[Pustilnik, Perepelitsuna \& Kniazev (2016)]{Pustilniketal16} and
\cite[Kniazev, Egorova \& Pustilnik (2018)]{Kniazevetal18}. Several new
void XMP galaxies are shown with green rombs.
A dozen void dwarfs  have $12+\log$(O/H) $< \sim$7.20, that is
$Z_{\rm gas} < \sim Z$\sunn/30. Of them, 4--5 dwarfs have O/H at or near
the so-called 'floor' level ($Z = Z$\sunn/50).
}
   \label{fig1}
\end{center}
\end{figure*}

\section{XMP dwarfs in the Lynx-Cancer, Eridanus and other nearby voids}

Here we briefly summarize our recently published results on the unbiased
study of Nearby Void galaxies
(\cite[Pustilnik, et al. (2010)]{Pustilniketal10},
\cite[Pustilnik \& Tepliakova (2011)]{Pustilnik2011},
\cite[Chengalur \& Pustilnik (2013)]{Chengalur13},
\cite[Perepelitsyna, Pustilnik \& Kniazev (2014)]{Perepeletal14},
\cite[Pustilnik \& Martin (2016)]{PustilnikMartin16},
\cite[Pustilnik, Perepelitsuna \& Kniazev (2016)]{Pustilniketal16},
\cite[Chengalur, Pustilnik \& Egorova(2017)]{Chengalur17},
\cite[Kniazev, Egorova \& Pustilnik (2018)]{Kniazevetal18})
or results from the papers in preparation.

1. Late-type void galaxies have in general reduced gas metallicity and
elevated gas content, both in average by $\sim$40\% with respect of reference
samples of similar galaxies in the denser environment of the Local Volume.

2. In voids all dwarfs with $M_{\rm B} > -13$ and with measured O/H, appear
   to be in the low metallicity regime (O/H $<$ (O/H)\sunn/10).

3. A dozen void dwarfs  have $12+\log$(O/H) $< \sim$7.20, that is
$Z_{\rm gas} < \sim Z$\sunn/30. Of them, 4-5 dwarfs have O/H at or near
the so-called 'floor' level ($Z = Z$\sunn/50).

4. The most gas-rich void dwarfs have M(HI)/$L_{\rm B} = 6-26$, or
$M_{*}$/$M_{\rm gas} \lesssim 0.01 - 0.02$.

5. Blue colours in the outer parts outside the Star-Forming (SF) regions
indicate relatively young stellar populations. Its time since the onset of SF
$t_{\rm SF} < \sim 1-3$ Gyr for several most extreme XMP dwarfs.  Their
baryon masses (consisting practically of gas) fall in the range of
$\sim 10^7$ M\sunn\  to $3\times10^8$ M\sunn.

\section{Diversity of XMP dwarfs and prospects of deeper insights}

There exist several scenarios explaining the existence of the most metal-poor
dwarfs. They include:

A) probably the most common scenario of the substantial metal loss via the
enriched gas outflow (wind) caused by SNe explosions during the episodes of
elevated SF.  One of the best examples is the nearest XMP dIr Leo~P with
$12+\log$(O/H) = 7.17. Its carefully studied SF history based on the deep HST
photometry, along with all other available data, indicates the loss of
$\sim$95\% metals produced during its cosmological evolution
(\cite[McQuinn et al. (2015)]{McQuinn2015});

B) temporary strong local dilution of metals in the regions of current SF
by blobs of the ambient intergalactic medium due to so called 'Cold Accretion'
along cosmological filaments of low-Z gas ($Z \sim Z$\sunn/50) (many examples
by Sanchez Almeida \& co-authors). However,
\cite[Filho et al. (2015)]{Filho2015}, and
\cite[Sanchez Almeida et al. (2016)]{Sanchez16})
also find that the low-metallicity dwarfs favor voids.

C) Inflow of very metal-poor unprocessed ($Z \gtrsim Z$\sunn/50) gas from the
distant periphery of gas-rich  dwarfs to the central region of straburst due
to the induced loss of gas stability by an external disturber
(\cite[Ekta \& Chengalur (2010)]{Ekta2010}).

D) Additional effect of void environment: slower evolution due to the
significantly reduced rate of galaxy interactions and probable delayed dwarf
formation in shallow gravitational potential of negative density contrast.
Such objects are expected to be true Very Young Galaxies (VYG, as defined by
Tweed et al. 2018). The SF galaxy J0811+4730 at $D = 180$ Mpc with
$Z_{\rm gas} = Z$\sunn/50 (Izotov et al. 2018) with only young stellar
population is a probable VYG. The most interesting candidates to such nearby
VYGs with lower SFR are presented here. A couple dozen similar candidates in
our Nearby Void Galaxy sample (Pustilnik et al. 2018, MNRAS, submitted) await
for the additional checks.

\begin{figure*}[b]
\begin{center}
 \includegraphics[width=12cm,angle=-0,clip=]{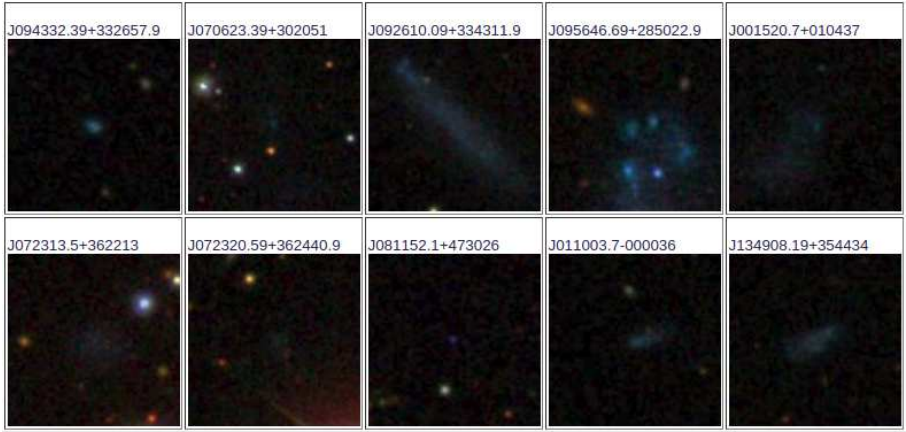}
 \caption{Eight Nearby Void dwarfs (including SF dwarf J0811+4730 at z=0.01444
from Izotov et al. 2018) with the confident or tentative $12+\log$(O/H) in the
range $\sim$7.0 -- 7.2, that is $Z_{\rm gas} \sim Z$\sunn/50 -- $Z$\sunn/30.
Two similar extremely gas-rich faint void dwarfs without O/H, J0723+3622 and
J0723+3624 are also included. Each finding chart is $\sim$50$''$ on side.}
   \label{fig2}
\end{center}
\end{figure*}

The work is supported by grant of Russian  Foundation for Basic Research
No.~18-52-45008.

\end{document}